\documentclass[
a4paper, superscriptaddress, twocolumn, amsmath, amssymb, amsfonts,frontmatterverbose, floatfix, aps,nofootinbib, 10pt%,rmp
]{revtex4-2}

\usepackage[english]{babel}
\usepackage[utf8]{inputenc}
\usepackage{graphicx}% Include figure files
\usepackage{dcolumn}% Align table columns on decimal point
\usepackage{bm}% bold math
\usepackage{physics}
\usepackage[dvipsnames,svgnames,x11names,hyperref]{xcolor}

\definecolor{bright}{RGB}{219, 48, 122}
\definecolor{dark}{RGB}{153, 34, 111}

\definecolor{bright}{RGB}{255, 90, 95}
\definecolor{dark}{RGB}{179, 63, 69}

\definecolor{bright}{RGB}{170, 68, 101}
\definecolor{dark}{RGB}{119, 48, 85}

\definecolor{bright}{RGB}{204, 41, 54}
\definecolor{dark}{RGB}{143, 29, 44}

\usepackage[pdftex]{hyperref}
\hypersetup{
    colorlinks=true,
    citecolor=bright,%WildStrawberry,%RedViolet,
    linkcolor=bright,%WildStrawberry,%OrangeRed,%WildStrawberry,%cyan,
    filecolor=bright,%WildStrawberry,      
    urlcolor=dark,%RedViolet,
    pdftitle={Resolving Photon Numbers Using Ultra-High-Resolution Timing of a Single Conventional Superconducting Nanowire Detector},
    }

\usepackage[left=17mm,right=17mm,top=18mm,bottom=28mm,columnsep=8mm]{geometry}

\begin{document}

\title{Resolving Photon Numbers Using Ultra-High-Resolution Timing of a Single Low-Jitter Superconducting Nanowire Detector}% Force line breaks with \\

\author{Gregor Sauer}
\email{gregor.sauer@uni-jena.de}%[Correspondence email address: ]
\affiliation{Friedrich Schiller University Jena, Fürstengraben~1, 07743 Jena, Germany}
\affiliation{Fraunhofer Institute for Applied Optics and Precision Engineering, Albert-Einstein-Str.~7, 07745 Jena, Germany}
\author{Mirco Kolarczik}
\email{mirco@swabianinstruments.com}
\affiliation{Swabian Instruments GmbH, Stammheimer Str.~41, 70435 Stuttgart, Germany}
\author{Rodrigo Gomez}%
\affiliation{Friedrich Schiller University Jena, Fürstengraben~1, 07743 Jena, Germany}
\affiliation{Fraunhofer Institute for Applied Optics and Precision Engineering, Albert-Einstein-Str.~7, 07745 Jena, Germany}
\author{Johanna Conrad}%
\affiliation{Friedrich Schiller University Jena, Fürstengraben~1, 07743 Jena, Germany}
\affiliation{Fraunhofer Institute for Applied Optics and Precision Engineering, Albert-Einstein-Str.~7, 07745 Jena, Germany}
%\author{Helmut Fedder}
%\affiliation{Swabian Instruments GmbH, Stammheimer Str.~41, 70435 Stuttgart, Germany}
\author{Fabian Steinlechner}%
\email{fabian.steinlechner@uni-jena.de}
\affiliation{Friedrich Schiller University Jena, Fürstengraben~1, 07743 Jena, Germany}
\affiliation{Fraunhofer Institute for Applied Optics and Precision Engineering, Albert-Einstein-Str.~7, 07745 Jena, Germany}
%\affiliation{Abbe Center of Photonics, Friedrich Schiller University Jena, Albert-Einstein-Str.~6, 07745 Jena, Germany}%
\date{\today}

\begin{abstract}
%\noindent
Photon-number-resolving (PNR) detectors are a key enabling technology in photonic quantum information processing.
Here, we demonstrate the PNR capacity of conventional superconducting nanowire single-photon detectors by performing ultra-high-resolution time-tagging 
of the detector-generated electrical pulses. 
This method provides a viable approach for PNR with high detection efficiency and a high operational repetition rate.
We present the implementation of such a PNR detector in the telecom C-band and its characterization by measuring the photon-number statistics of coherent light with tunable intensity. Additionally, we demonstrate the capabilities of the detection method by measuring photon-number correlations of non-classical states.

\end{abstract}

\maketitle

\section{Introduction}
\label{sec:intro}
Efficient detectors play a crucial role in photonic quantum technology, and many recent milestones in the field can be traced back to the advancement of detector technology over the past decades \cite{CHRIST2013351, marsili2013detecting, zwiller_high_QE_low_jitter}. Two types of detectors can be distinguished: threshold detectors, which provide a "click" in presence of one or more photons, and detectors with photon number resolution (PNR).

PNR is an essential functionality for protocols like quantum teleportation and entanglement swapping \cite{Zukowski_1993_EventreadydetectorsBellExperiment}, photonic quantum computing \cite{Kok_2007_LinearOpticalQuantum}, and quantum-enhanced sensing \cite{Edamatsu_2002_MeasurementPhotonicBroglie}. Transition-edge sensors (TES) can be used, which intrinsically provide energy- and photon-number resolution encoded in the pulse form of the detector response \cite{rosenberg_nam_pnr_tes, Lita_2008_CountingNearinfraredSinglephotons, nam_TES_PNR, schmidt2018photon}. This comes at the cost of long recovery times. 
Presently, a commonly employed alternative approach is to achieve PNR functionality by multiplexing the photon flux onto several threshold detectors (e.g.~superconducting nanowire single-photon detectors (SNSPDs)), where spatial-, temporal-, and multi-pixel approaches have been demonstrated  \cite{Jonsson_2020_PhotoncountingDistributionArrays, achilles, divochiy, Fitch_2003_PhotonnumberResolutionUsing, stasi2022high}. 
This approaches are already established, but continuous further development in terms of efficiency and repetition rate are a key challenge. 

Recent studies \citep{Cahall:17, Endo:21}, however, have provided findings that challenged the conventional perception of SNSPDs as threshold detectors. These studies demonstrated that the waveform of SNSPD detector pulses can provide information about the number of photons absorbed within the thermal relaxation time of the nanowire \cite{Natarajan_2012}. In a comprehensive detector tomography of an SNSPD, \citet{Endo:21} showed that the photon number can be inferred by matching waveform traces recorded in a high-resolution oscilloscope. Moreover, the turn-on dynamics of SNSPDs, as modeled by \citet{Nicolich_2019_UniversalModelTurnOn} indicate that the rising slope of that waveform increases with a higher photon number of a detection event. Correspondingly, an earlier trigger at a particular threshold value should also allow for a distinction between photon numbers. This suggests that PNR detection should be possible using time-tagging devices with exceptional timing resolution.

In this work, we empirically study the timing statistics of SNSPD detector waveforms for photon wavepackets with different photon number distributions using ultra-fast time-tagging devices. Our experimental findings reveal that for our detection system and amplification circuit, simple threshold detection of the rising edge offers basic photon number resolution (PNR) functionality up to approximately N=2. Moreover, we find that substantial improvement of the PNR functionality can be achieved by not only considering the rising edge but also classifying detection events by the timing of the falling edge of the SNSPD-generated electrical pulse. This is also in good agreement with recent findings from an extensive principle component analysis of SNSPD electrical pulse shapes \citep{bartley}.

We experimentally implement this PNR detection scheme in the telecom C-band and characterize it by measuring the photon-number statistics of a $0.3$ ps-pulsed coherent light source with a tunable average photon number. We demonstrate that high-resolution time-tagging of the rising and falling edge of pulses enables PNR measurements with a single conventional low-jitter SNSPD, resolving up to N=5 photons in a single-shot measurement. To further demonstrate the applicability to quantum applications, we employ the approach for measurements of the joint photon-number distributions (JPND) of N00N states (N=2) generated from type-II spontaneous parametric down-conversion (SPDC) in a Hong-Ou-Mandel interferometer.

The manuscript is organized as follows. In Section~\ref{sec:detection_scheme}, we explain the detection approach and then evaluate the performance of the detection scheme using coherent states of light with different mean photon numbers. Further, we demonstrate the effectiveness of the method by measuring joint photon-number distributions of non-classical states in Section~\ref{sec:n00n}. The main findings are summarized in Section~\ref{sec:conclusion}, together with an outlook on future steps towards broader impact.

\renewcommand{\thefigure}{1}
\begin{figure*}
\centering
\includegraphics[width=0.7\linewidth]{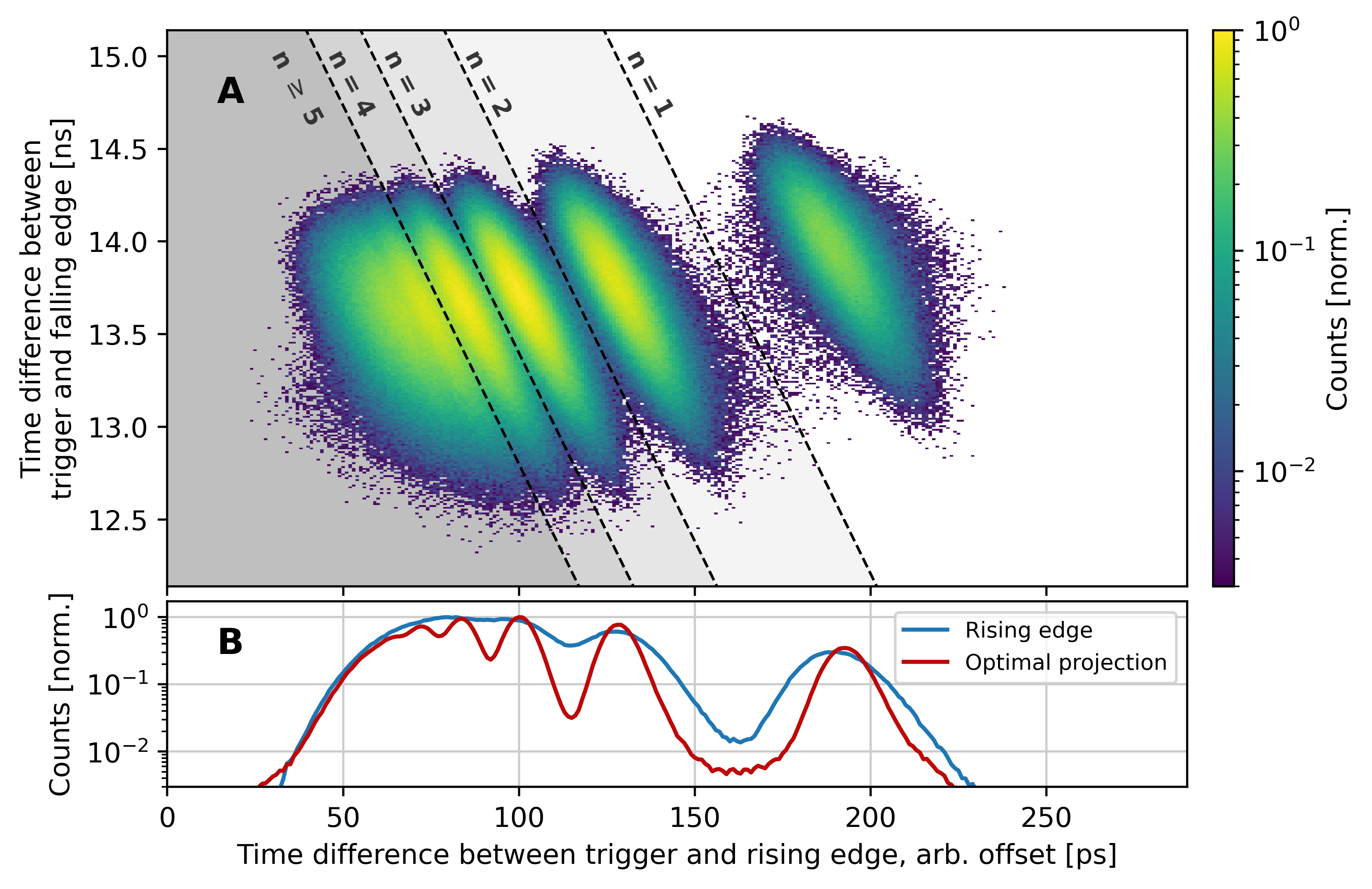}
\caption{\textbf{A}: A two-dimensional histogram of the time difference between the trigger of a pulsed laser and the rising and falling edges of electrical pulses generated from detection events. The histogram shows well-distinguishable modes that can be related to the photon numbers. We verify this by correctly predicting the photon number statistics of various $\mu$ (see main text). \textbf{B}: Projections of the histogram to the rising edge (blue) and for optimal mode distinction (red, along mode separation boundaries in Fig.~\ref{fig:blobs}~A). Also considering the falling edge results in substantially better distinguishability of photon-number modes. }
\label{fig:blobs}
\end{figure*}

\section{Results}
\subsection{PNR Characterization using coherent states}
\label{sec:detection_scheme}

\subsubsection*{Experimental Setup}
\label{sec:setup}

The experimental setup for PNR detection is shown in Fig.~\ref{fig:setup}. We produce coherent light pulses with low mean photon numbers $\mu$ by attenuating laser pulses at $1554$ nm with a duration of $0.3$ ps. To avoid saturation of the SNSPD we reduce the repetition rate of the laser system from $400$ MHz to $10-500$ kHz via pulse picking. For count rates up to approx.~$1$ Mcps, the detection efficiency of the SNSPD (Single Quantum, EOS Series) is specified as $86 \%$. 

For each measurement run, the detection efficiency is optimized by alignment of the polarization of the light with a fiber polarization controller placed before the SNSPD.

\renewcommand{\thefigure}{2}
\begin{figure}[ht]
\centering
\includegraphics[width=0.9\linewidth]{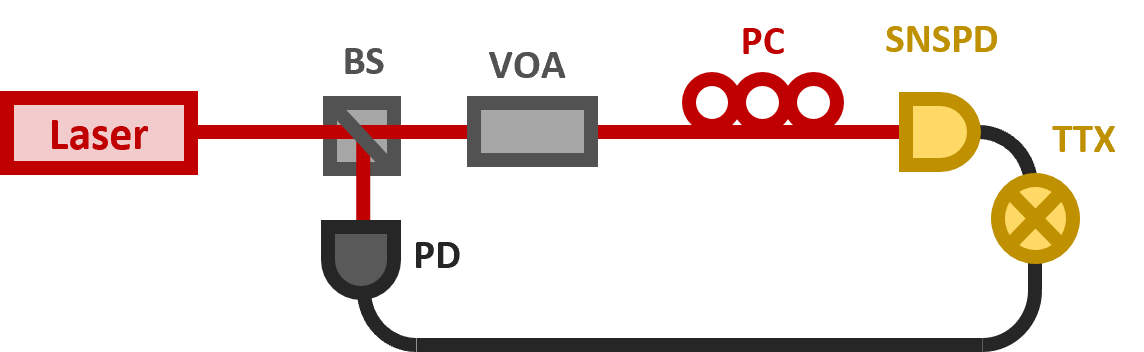}
\caption{Experimental setup for the characterization of the PNR detection method with attenuated laser pulses.\\ BS: beam splitter, PD: photodiode, VOA: variable optical attenuator, PC: polarization controller, SNSPD: superconducting nanowire single-photon detector, TTX: ultra-high-resolution time-tagging unit.}
\label{fig:setup}
\end{figure}

The arrival time of the wave packet on the SNSPD is precisely determined by splitting off a part of the laser with a beam splitter and detecting it with a high-bandwidth photodiode. The electric pulses from the photodiode and the SNSPD are then converted to time stamps using an ultra-high-resolution time-tagging unit (TTX, Swabian Instruments, TimeTagger X). 

\subsubsection*{Results and Discussion}
Fig.~\ref{fig:blobs}~A shows the two-dimensional histogram of the difference between the arrival time of the wave packet on the detector and the timing of the rising and falling edge  of the electrical pulses of the SNSPD. 
The histogram exhibits distinct modes that correspond to the photon number in the coherent state, with a detected mean photon number of $\mu = 3.43$.

In order to determine the photon number of individual detection events, a simple analysis of the time difference between the trigger and the rising edge (which corresponds to projection along vertical lines in Fig.~\ref{fig:blobs} A) allows to distinguish between N=1 and N\textgreater 1 with high accuracy. Essentially, this can be understood as a consequence of the turn-on dynamics and the steeper rising edge of the detection waveform \cite{Nicolich_2019_UniversalModelTurnOn}. 
It is, however, evident that a clearer distinction of the photon-number modes can be found by considering not only the rising edge, but also the falling edge to optimally cluster detection events. A simple ad-hoc approach is to separate the clusters with parallel lines at an angle, i.e., by comparing different projections of the 2D histogram, along the falling edge timing and the found distinction lines. The optimal position and angle for the separation can be found by numerical optimization, minimizing the cross-talk between photon-number modes. A comparison in Fig.~\ref{fig:blobs}~B shows the significantly improved distinction between photon-number modes enabled by consideration of the falling edge timing. 
Zero-photon detection events are inferred from photodiode detections events without corresponding detection events on the SNSPD.

Note that this clustering of detector modes is only possible with low timing jitter of the measurement. The main contributors to the measured jitter is the SNSPD with a low jitter of $8.1$ ps root mean squared (RMS), enabled by cryogenic amplification, and the TTX with a jitter of $1.3$ ps RMS per channel.

To confirm the correspondence between clusters and  photon numbers, we characterize the photon number statistics of coherent states of light. We vary the mean photon number of the coherent light by means of attenuation and compare the measured photon-number statistics with the expected Poissonian photon-number distributions.

\renewcommand{\thefigure}{3}
\begin{figure}[hb]
\centering
\includegraphics[width=0.85\linewidth]{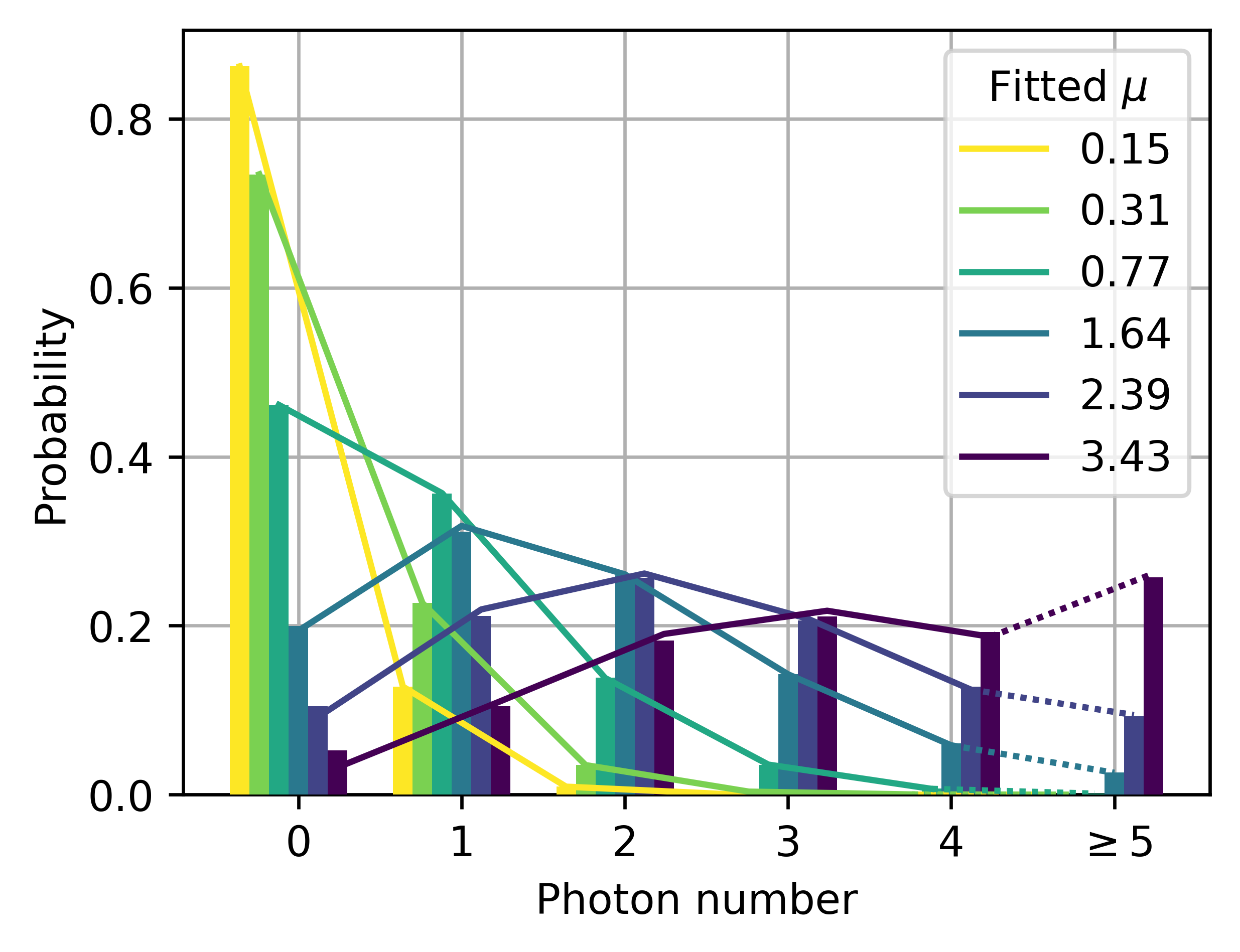}
\caption{The measured photon-number statistics (bar plot) of coherent light at different attenuation levels show good agreement with the fitted Poissonian photon-number distribution (solid line). The deviation from Poissonian behaviour is due to truncation at N=5 (dotted line). Mean photon numbers $\mu$ were obtained by a fit.}
\label{fig:coherents}
\end{figure}

The measured photon-number statistics shown in Fig.~\ref{fig:coherents} are in good agreement with the fitted Poissonian photon-number distributions (solid line, truncated at N=5). Values for $\mu$ were obtained by fitting the truncated photon-number distribution (summing up occurrences of photon numbers N$\geq$4). Graphical inspection shows the clear distinguishability of photon-number modes up to N=4.

To better quantify the benefit of optimal clustering, we model the marginal timing distribution (along the optimal projection axis) as a sum of the individual cluster distributions. We observe excellent agreement with our experimental data when modeling the individual clusters using a Voigt profile fit, as shown in Fig.~\ref{fig:voigt}. %For further details refer to the supplementary material. 

\renewcommand{\thefigure}{4}
\begin{figure}[h]
\centering
\includegraphics[width=1.0\linewidth]{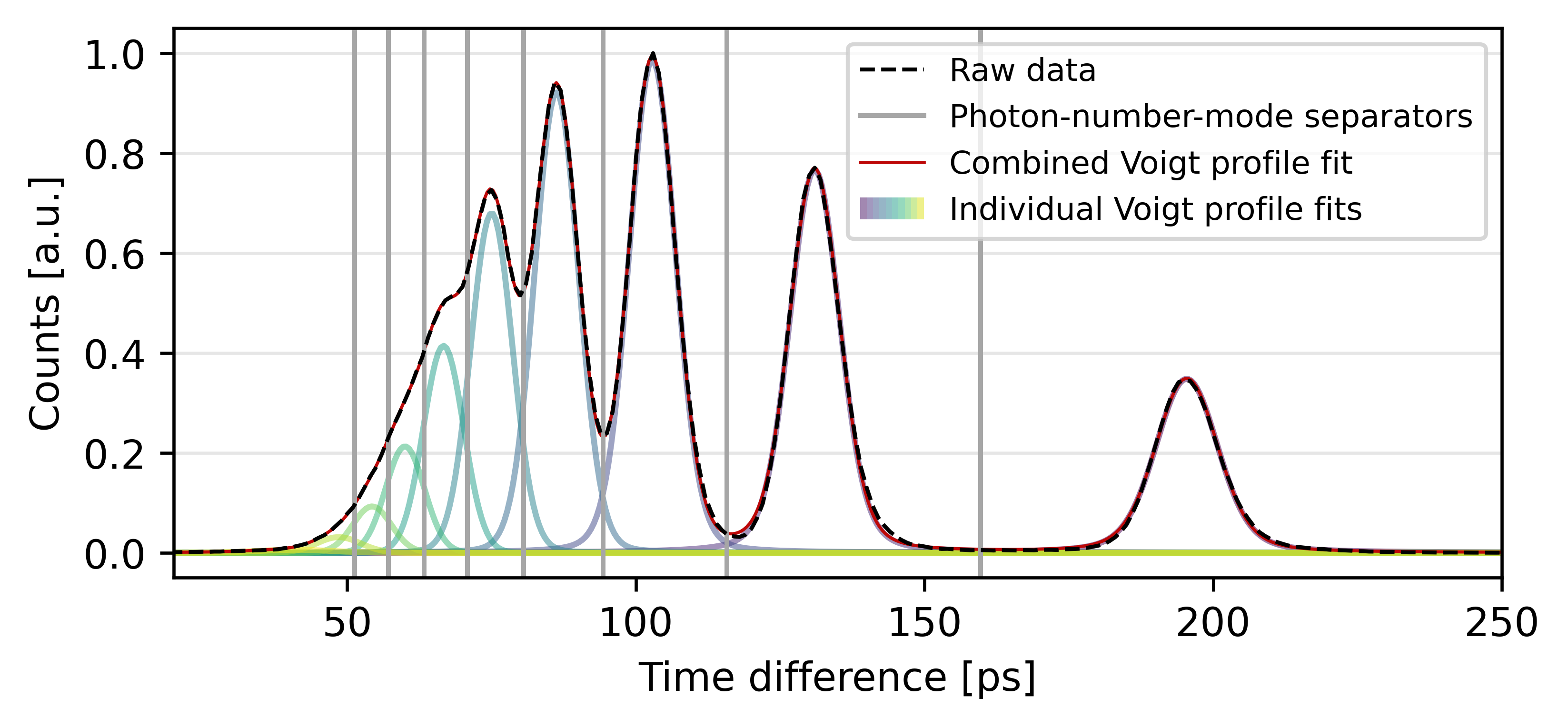}
\caption{Histogram of the time difference between the trigger of a pulsed laser and the rising and falling edges of the SNSPD signal with optimal clustering (see Fig.~\ref{fig:blobs} B). Voigt profiles are fitted to the histogram and show excellent agreement with the measured data.}
\label{fig:voigt}
\end{figure}

By calculating the overlap between the fitted Voigt profiles and the identified photon-number bins, we can determine the cross talk between the photon-number channels. Fig.~\ref{fig:crosstalk} displays the cross talk calculated for photon-number measurements conducted using only the rising edge timing and optimal clustering methods. Comparing these cross talk matrices reveals a significant reduction in cross talk when utilizing the optimal clustering method. This reduction enables a more precise identification of photon numbers in measurements.

\renewcommand{\thefigure}{5}
\begin{figure}[h]
\centering
\includegraphics[width=1.0\linewidth]{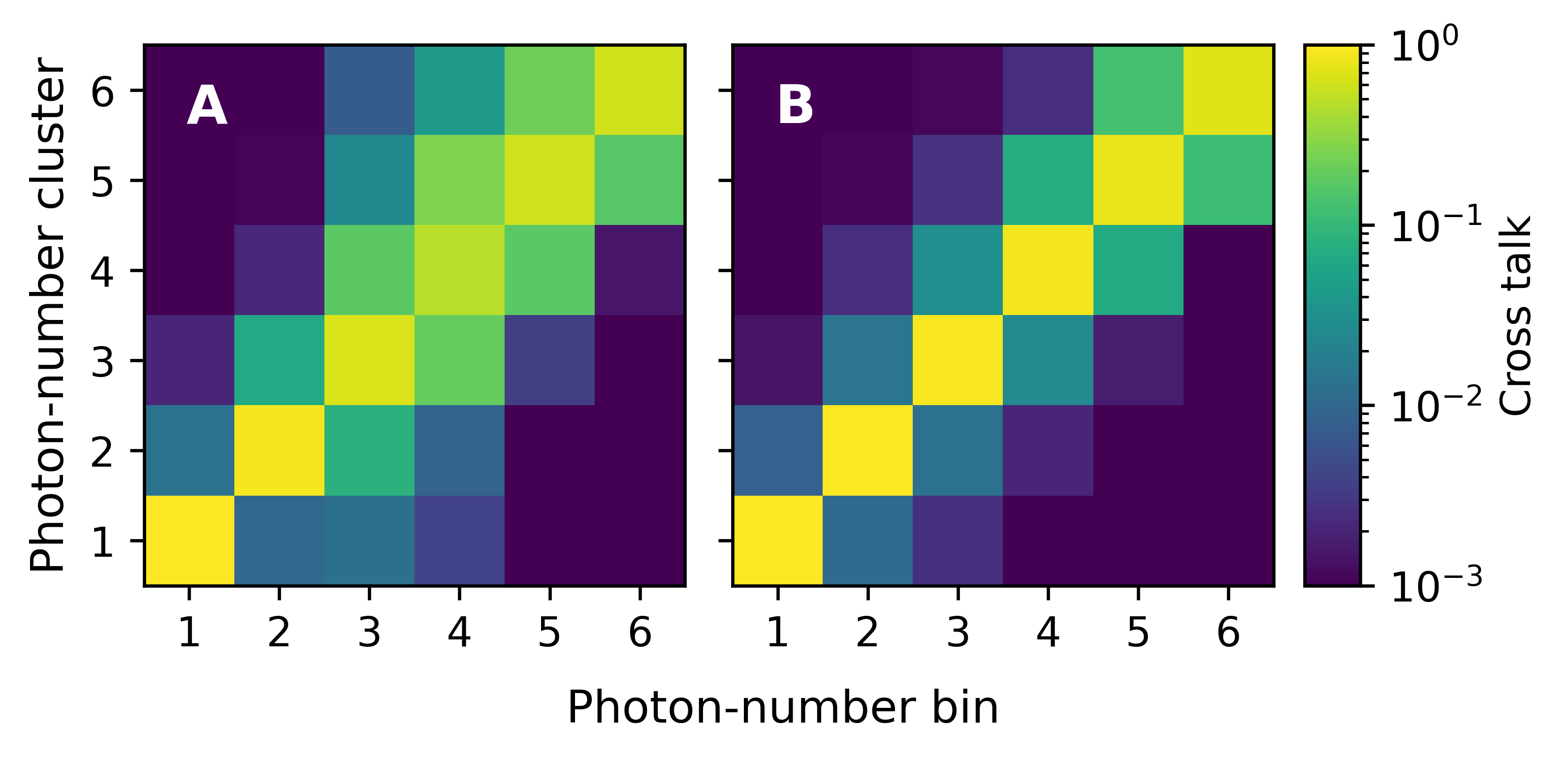}
\caption{Cross talk between photon-number clusters is calculated using fitted Voigt profiles for PNR measurements. The comparison between matrices from two different methods (\textbf{A}: rising edge timing and \textbf{B}: optimal clustering) reveals that considering the falling edge timing significantly reduces cross talk between the photon-number channels.}
\label{fig:crosstalk}
\end{figure}

\subsection{Joint PNR Detection using non-classical light}
\label{sec:n00n}

\renewcommand{\thefigure}{6}
\begin{figure*}%[b!]%[ht]
\centering
\includegraphics[width=0.65\linewidth]{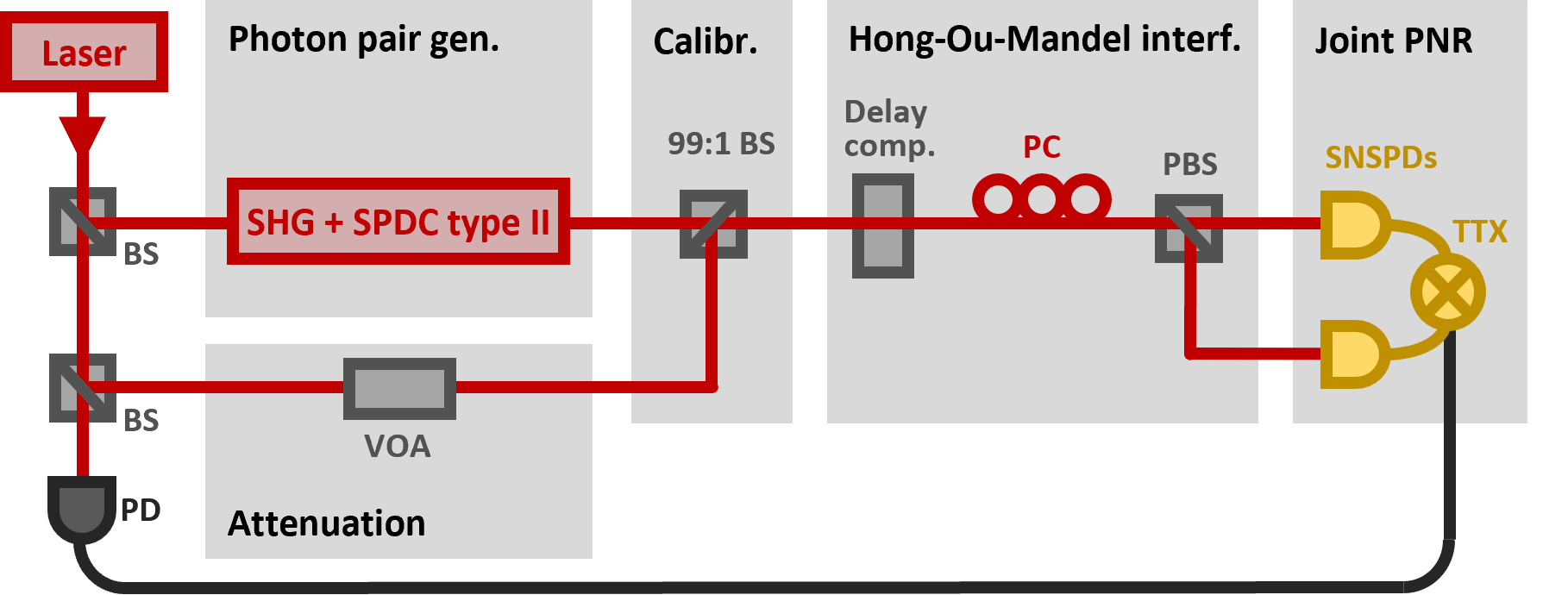}
\caption{Experimental setup for the generation of N00N states and the characterization of their joint photon-number distributions by correlated PNR measurements. BS: beam splitter, PD: photodiode, VOA: variable optical attenuator, PBS: polarizing beam splitter, PC: polarization controller, SNSPD: superconducting nanowire single-photon detector, TTX: time-tagging unit.}
\label{fig:setup_n00n}
\end{figure*}

\subsubsection*{Experimental Setup}
%The utility of the shot-by-shot PNRD is further exemplified by using the detection scheme to characterize the joint photon-number distribution of two-mode N00N states. 
To generate and characterize the joint photon-number distribution of N00N states \cite{dowling2008quantum}, the experimental setup was modified as shown in Fig.~\ref{fig:setup_n00n}. The N00N states with N=2 are generated by interfering photon-pairs from a type-II SPDC process in a Hong-Ou-Mandel interferometer. The photon pairs are generated in a periodically poled potassium titanyl phosphate (ppKTP) crystal, pumped by the second harmonic of the pump laser at $1554$ nm. The two-step non-linear process to generate the photon pairs is pumped with a pulsed laser with a repetition rate of $400$ MHz and a transform-limited pulse duration of approx.~$1.2$ ps. In this experiment, no pulse picking was required due to the low average count rates produced by the SPDC process. Every detection event resulting in a photon number measurement is associated with a time tag. Thus, we can identify correlated detection events across several detectors, enabling the characterization of joint photon-number distributions.

The transformation of the polarization state in the common-path Hong-Ou-Mandel interferometer allows switching between creating N00N states, and deterministically splitting photon pairs created in the SPDC process. The generation of N00N states is achieved by minimizing coincidence counts between the two SNSPD channels. Respectively, a maximization of those coincidence counts results in the deterministic splitting of photon pairs generated in the SPDC process.

The second detector channel employed for these measurements has an estimated detection probability of $91 \%$ and a timing jitter of $9.2$ ps RMS.

\subsubsection*{Results and Discussion}

The obtained JPNDs in Fig.~\ref{fig:matrices} show the expected photon-number correlations between the two PNR detectors depending on the polarization transformation in the Hong-Ou-Mandel interferometer. The difference between the two generated states becomes especially evident when one examines the occurrences of two-photon events depicted in Fig.~\ref{fig:bars}. This confirms the accurate identification of correlations of individual PNR measurements. 

\renewcommand{\thefigure}{7}
\begin{figure}[ht]
\centering
\includegraphics[width=1.0\linewidth]{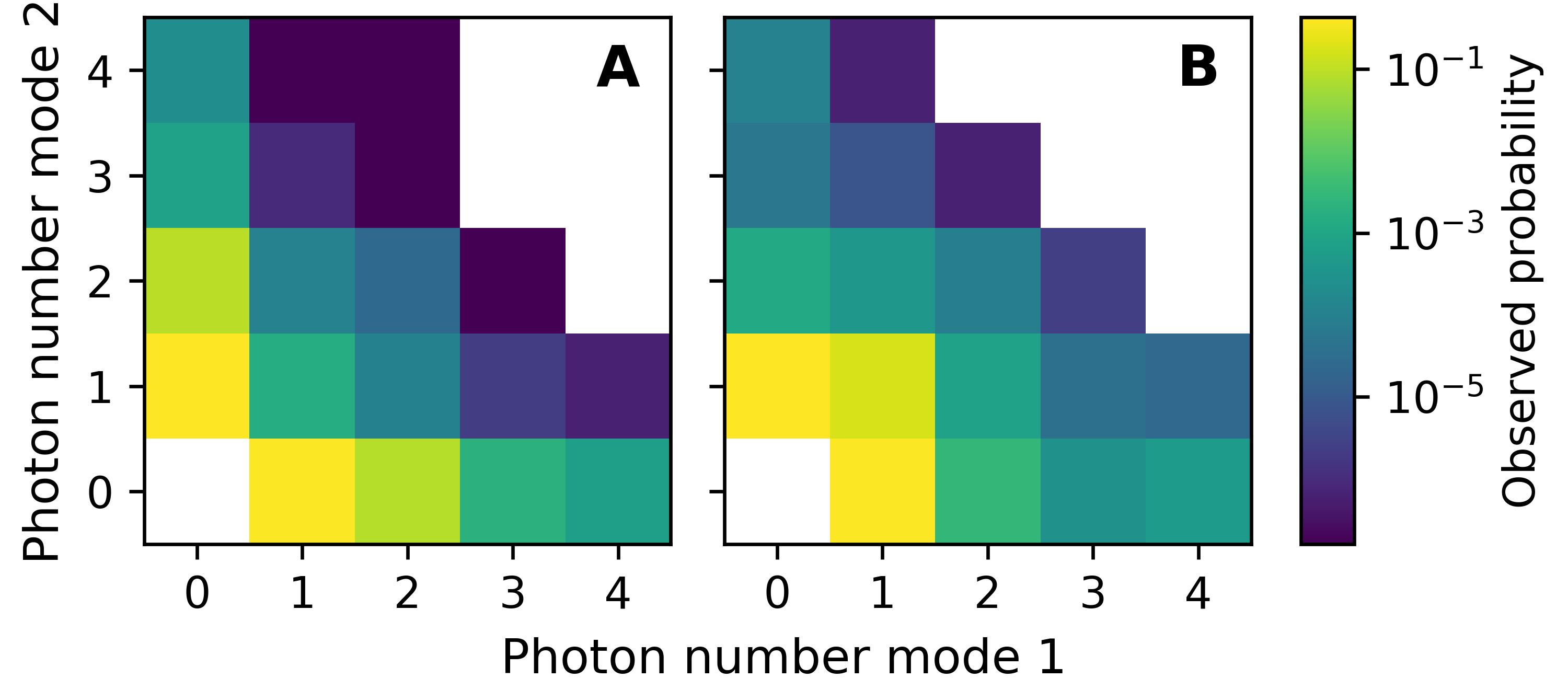}
\caption{Joint photon-number distributions measured for two different polarization settings. Measurements with two different prepared states: \textbf{A}  N00N states (N=2) and \textbf{B} split photon pairs.}
\label{fig:matrices}
\end{figure}

Furthermore, the observed JPNDs provide further information on the system's performance parameters. For example, by comparing the rates of detecting single photons with the rates of detecting photon pairs, it is possible to estimate the total system efficiency of approx.~$30\%$ (including fiber coupling efficiency, system transmission, and detection efficiency).  %a comparison of single-photon detection rates with photon pair detection rates enables the estimation of the total system efficiency of approx.~$30\%$. 
The extension to higher-order photon number contributions could provide further characterization metrics, such as multi-pair generation rates.

\renewcommand{\thefigure}{8}
\begin{figure}[ht]
\centering
\includegraphics[width=0.8\linewidth]{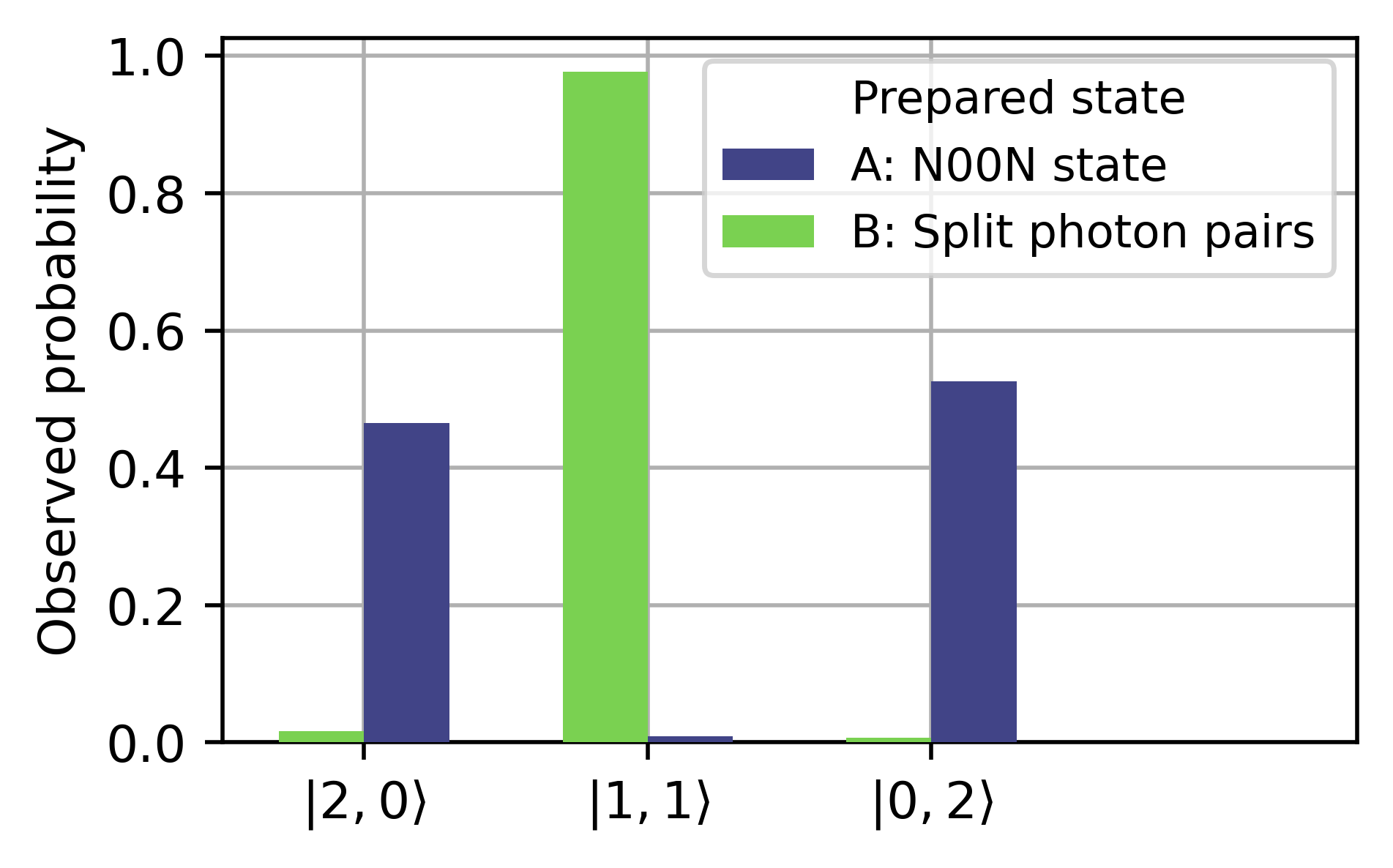}
\caption{Two-photon events from JPND measurements shown in Fig.~\ref{fig:matrices}. Differences in count-rates between $\ket{2,0}$ and $\ket{0,2}$ events result from different system efficiencies in the optical paths of the experiment.}
\label{fig:bars}
\end{figure}

\section{Conclusions}
\label{sec:conclusion}

In this work, we have explored the implementation of a PNR detector using a single conventional low-jitter superconducting nanowire detector and threshold analysis of the electric waveform with ultra-high-resolution timing. 

% Kurz nochmal mehrwert ggü etablierten Ansätzen erwähnen --> 
The method requires precise knowledge of the arrival time of the wave packet on the detector, either by splitting off a part of the wave packet or by analysis of a pump pulse involved in a non-linear optical process needed for the generation of a certain state. Nevertheless, this approach can dramatically reduce resource overhead compared to approaches based on multiplexing and 
promises substantially shorter recovery rates than PNR detectors based on TES.

We have shown a successful implementation and characterization of the detection method with clear distinguishability of photon modes up to $5$ photons. This is confirmed by showing good agreement of measured photon-number distributions of a coherent light source with Poissonian photon-number statistics. It also indicates a potential extension to a simple PNR detection tomography method using coherent states of light. Coherent states offer several advantages for the characterization of the PNR detection system. They are easily and reliably generated and offer non-vanishing probabilities for higher photon numbers, even at low intensities. Additionally, they are resilient to the inevitable losses in the light propagation and detection and thus offer a good reference state. However, this property makes it impossible to assess the loss characteristics of the detection scheme. The method thus needs to be employed with accurately characterized attenuators. Alternatively a well-characterized threshold detector could provide a necessary reference. A detailed discussion of absolute detector tomography, however, is beyond the scope of this initial work. 

Furthermore, we demonstrate the usefulness of the shot-by-shot PNR detection approach by characterizing the joint photon-number distributions of N00N states with N=2. This would be a daunting task for methods that rely on the analysis of the complete waveform, in particular when scaling to higher photon numbers.

In conclusion, we have demonstrated that it is possible to resolve high photon numbers with minimal overhead and infrastructure already available to researchers in many labs. We note that the approaches outlined herein are heavily dependent on the specifics of the detection system %, the amplification, 
and hope that our results will inspire further studies into the prediction of the turn-on and turn-off behaviour of superconducting nanowires. % and amplification circuits. %Despite remaining knowledge gaps in the description of the observed phenomena, 
Although a more comprehensive understanding of the observed phenomena needs to be explored in further work, 
we are confident that our empirical results already provide a simple means for leveraging PNR capability of existing detectors that are widely available in the research community. 

Moreover, we believe that further enhancement of the approach could be accomplished either by suitable pre-conditioning of electronic detector signals or by evaluating more than two threshold levels.

\section*{Acknowledgments}

We acknowledge support from the European Union’s Horizon 2020 Research and Innovation Action under Grant Agreement No.~899824 (SURQUID). GS acknowledges funding of this project from the Carl-Zeiss-Stiftung. 
GS is a member of the Max Planck School of Photonics, supported by the German Federal Ministry of Education and Research, the Max Planck Society, and the Fraunhofer Society. 
The authors gratefully acknowledge the support of René Sondenheimer, Oskar Kohout, Cristina Amaya, Shreya Gouravaram, and Helmut Fedder for their invaluable contributions to discussions and extend their thanks to Markus Leipe and Cristina Amaya, for their assistance in operating the SPDC source. 
We also acknowledge Single Quantum B.V.~for providing a preliminary experimental setup and their detector systems for first tests.

\bibliography{pnr}

%~\\

\end{document}